\documentclass[twocolumn,showpacs,preprintnumbers,amsmath,amssymb,aps,prb]{revtex4}

\usepackage{graphics}
\usepackage{color}
\begin{document}

\title{Magnetic order in coupled spin-half and spin-one Heisenberg chains in
anisotropic triangular-lattice geometry}%
\author{T.Pardini}
\author{R.R.P. Singh}%
\email[R.R.P Singh: ]{singh@raman.physics.ucdavis.edu}
\affiliation{Department of Physics, University of California, Davis, California 95616,USA}
\date{Jan 18,2008}%
\begin{abstract}
We study spin-half and spin-one Heisenberg models in the limit where one dimensional (1-D) linear chains, with exchange constant $J_1$,
are weakly coupled in an
anisotropic triangular lattice geometry. Results are obtained by means of linked-cluster series expansions at zero temperature around different 
magnetically ordered phases. We study the non-colinear spiral phases that arise classically in the model
and the colinear antiferromagnet that has been recently proposed for the spin-half model by Starykh and Balents using a Renormalization Group approach.
We find that such phases can be stabilized in the spin-half model for arbitrarily small coupling between the chains.
For vanishing coupling
between the chains the energy of each phase must approach that of
decoupled linear chains.
With increasing inter-chain coupling, the non-colinear
phase appears to have a lower energy in our calculations. 
For the spin-one chain, we find that there is a critical interchain coupling needed to
overcome the Haldane gap. When spin-one chains are coupled in an unfrustrated manner, the critical coupling 
is very small ($\approx 0.01 J_1$) and agrees
well with previous chain mean-field studies. When they are coupled in the
frustrated triangular-lattice geometry, the critical coupling required to 
develop magnetic order is substantially larger ($> 0.3J_1$).
The colinear phase is not obtained for the spin-one Heisenberg model.
\end{abstract}
\pacs{Valid PACS appear here}
\maketitle
\section{Introduction}
\begin{figure}[ht]\label{square}
\resizebox{40mm}{!}{\includegraphics{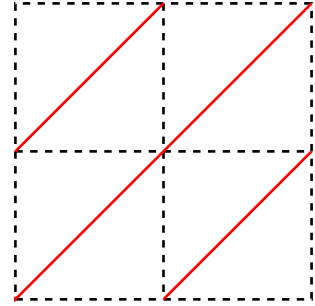}}
\caption{\label{fig:square}(Color online) Square lattice with coupling constant $J_2$ (dashed black line) along the horizontal and vertical axis and $J_1$ (solid red line) 
along the diagonal}
\label{test}
\end{figure}

There has been considerable recent interest in the properties of two-dimensional (2-D) antiferromagnetic Heisenberg models. 
In the absence of frustration, the ground state phases and properties of these models are quite well understood\cite{Auerbach1994,Sachdev1995,Chakravarty1988}. 
On the contrary, a complete knowledge of the ground state phase diagram of frustrated Heisenberg models is still lacking.\\
In the present paper, we further study the antiferromagnetic Heisenberg model on the
anisotropic triangular lattice\cite{Zheng1999}.
For this class of models, the Hamiltonian can also be defined on a square lattice with nearest neighbor interaction $J_2\geq 0$ 
and a second-neighbor interaction $J_1=1$ along one of the diagonals of the squares, as shown in Fig.~\ref{fig:square}. 
In the limit $J_2\rightarrow 0$, the model reduces to 1-D decoupled spin chains along 
the diagonals. 
In the limit $J_2=1$ the model becomes the Heisenberg model on the 2-D isotropic triangular lattice, for which there is
strong numerical evidence of long-range antiferromagnetic 
order\cite{Bernu1994,Singh1992,Zhong1993,capriotti1999}.
We are interested in studying the region $0<J_2<1$ to see how the order develops as the interchain
coupling is increased.

One reason for strong interest in these models is that
the Heisenberg model with $J_2\sim 1/3$ provides the dominant terms in the
Hamiltonian for the material Cs$_2$CuCl$_4$\cite{Coldea1997}.
The exchange parameters of this material have been determined from high-field studies\cite{Coldea2002,Coldea2003}, by
measuring the excitation energies around the fully polarized limit. 
In zero field, these materials are found to have spiral long-range order in the 
ground state\cite{Coldea1996b}. The materials also have non-zero Dzyloshinski-Moria (DM) interactions.
Spin-wave theories\cite{veillette2005,dalidovich2006} can
account for the spin-wave dispersion in the materials only after substantially modifying the
exchange parameters. On the other hand, series expansions around the spiral state give an excellent account of the main peaks 
observed experimentally in the spectra throughout the Brillouin zone, with the observed exchange
parameters\cite{restad2007,zheng2006,zheng2005}.
In another approach to the spectra of these materials,
Kohno et al.\cite{Kohno2007} have shown that many features of the experimental spectra, including the observed continuum can be well
explained by considering weakly coupled Heisenberg chains, even though the interchain coupling is not too small in the material. 

In another recent study, directly relevant to the present work,
Starykh and Balents\cite{starykh2007} have considered the frustrated weakly coupled chain problem using a 
Renormalization Group approach. Their striking prediction is
that a colinear-phase is stabilized in place of the classical spiral phase for sufficiently weak coupling
between the chains.
The phase corresponds precisely to the four-sublattice phase that is known to
occur in the square-lattice $J_1-J_2$ model at large $J_2$\cite{schulz,oitmaa1996}. Other
analytical, numerical and variational studies have also been used to study this
anisotropic triangular-lattice model\cite{alicea2005,alicea2006,yunoki2006,Chung2001,Merino1999}.
Several of these have found evidence for disordered spin-liquid phases in the model
at weak coupling between the chains. 
Most notably the variational calculations of Sorella and coworkers find two distinct spin-liquid phases in the model.
Furthermore, the DMRG studies\cite{Weng2006} of Weng \textit{et al.} showed a very rapid exponential
decrease in correlations perpendicular to the chains even for
rather large interchain couplings. One possible concern with these studies is that 
periodic boundary conditions were used and that can play a role in destabilizing non-colinear phases if they occur at
incommensurate wavevectors.  Earlier series expansion studies\cite{Zheng1999} also found
that the energies from spiral-phase series expansions and dimer
expansions were very close. In general, numerical studies of weakly coupled chains in frustrated geometries have been a challenging problem\cite{moukouri2004,arlego2007}. 
 
In this paper, we revisit this model for the spin-half case and
also study the corresponding spin-one model.
To our knowledge, this is the first quantitative 
numerical study of the spin-one model. 
We use an Ising-type linked-cluster expansion method\cite{gelfand2000} at zero 
temperature around different phases. Ground state energy and sublattice magnetization have been calculated for each phase.
The knowledge of the exact behavior of the 1-D spin-half model\cite{bethe1931}, with its essential 
singularity in the energy and magnetization functions\cite{baxter1973}, 
is used to improve the series extrapolations in the 1-D limit, and they are also used to get
more accurate estimates of the ground state energies with inter-chain
couplings.
While both colinear and non-colinear phases can be stabilized in our studies for the spin-half model, we always find the energy of the non-colinear
phase to be lower. However, the results are quite sensitive to the way the series
are analyzed, especially in the limit of weak interachain couplings,
and this implies some uncertainties in our results
that cannot be addressed by series expansions alone. Further
numerical studies of these models would be useful.

We have also investigated the spin-one model, with unfrustrated (square-lattice geometry) and
frustrated (triangular-lattice geometry) interchain couplings. In this case, the 1-D limit corresponds to a Haldane gap phase.
The Ising expansions are known to break down before the Heisenberg symmetry is restored, with a critical point which is in the
universality class of the 2-D Ising model. We find that when the chains are coupled in an unfrustrated manner a
rather small interchain coupling ($J_2/J_1<0.01$) leads to Ne\'{e}l order. On the other hand, in the frustrated geometry
a much larger interchain coupling ($J_2/J_1>0.3$) is needed to obtain long-range order. In the latter case, we only
find the spiral phase to be stabilized for Heisenberg models. The colinear-phase becomes less and less stable with
increased inter-chain coupling. It should be stressed that the analysis of Starykh and Balents\cite{starykh2007} was special for the
spin-half case and hence there is no apriori reason to expect a colinear phase in the spin-one model. 

Our results are organized as follows. In Section~\ref{seriesexpansion} we discuss the methods of series expansions. In Section~\ref{1Dspinhalf} 
ground state energy and sublattice magnetization for the spin-half chain are presented. In Section~\ref{2Dspinhalf} 
we discuss our results for the spin-half model on the anisotropic triangular lattice. In Section~\ref{spinone} 
the study of the spin one model is presented. Finally, in Section~\ref{conclusions} we present our conclusions.

\section{Series expansion}\label{seriesexpansion}
\begin{figure}
  \resizebox{55mm}{!}{\includegraphics{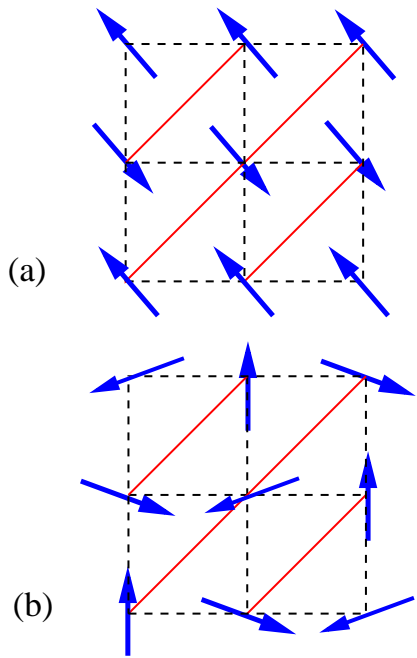}}
  \caption{\label{fig:caf}(Color online) (a) CAF phase: spins are aligned antiferromagnetically along the diagonals and the vertical axis of the square and ferromagnetically along the horizontal axis.
(b) Classical spiral phase.}
\end{figure}

The antiferromagnetic Heisenberg model is defined by the Hamiltonian
\begin{equation}\label{Hamiltonian}
H=J_1\sum_{[i,j]}\textbf{S}_i\cdot \textbf{S}_j+J_2\sum_{<i,j>}\textbf{S}_i\cdot \textbf{S}_j.
\end{equation}
Here, $[i,j]$ refers to one of the diagonal next-nearest-neighbor pairs on the square-lattice shown in Fig.~1,
with corresponding coupling constant $J_1$, while 
$<i,j>$ are pairs of nearest-neighbors with coupling constant $J_2$. 
We set $J_1=1$, and vary $J_2$ in the range $0\leq J_2\leq 1$. 
In the limit $J_2=0$ the model is equivalent to decoupled antiferromagnetic spin chains and is exactly solvable
for spin-half\cite{bethe1931}. 
For $J_2=1$, the model is equivalent to the Heisenberg model on an isotropic triangular lattice. In this limit the classical ground state has
a $3$-sublattice `120-degree' order, which can also be regarded as a
non colinear spiral with wave wector $q=\cos^{-1}(-\frac{1}{2})=\frac{2\pi}{3}$. 
The predicted CAF phase of Starykh and Balents\cite{starykh2007} for small $J_2$
is shown in Fig.~\ref{fig:caf}a. In this phase, the spins are aligned antiferromagnetically along the diagonals and the vertical axis of the square lattice, 
and ferromagnetically along the horizontal axis.
A sketch of the classical spiral phase is shown in Fig.~\ref{fig:caf}b.
It was found in earlier series expansion studies\cite{Zheng1999} that
away from the triangular lattice limit, quantum fluctuations renormalize the angle $q$ with respect to the classical value.
We will refer to this renormalized spiral phase as the non colinear antiferromagntic 
phase (NCAF).

To obtain a $T=0$ expansion about the CAF phase, the Hamiltonian is written as
\begin{equation}\label{Hamiltonianform}
  H=H_0+\lambda (H_1+H_2)
\end{equation}
where
\begin{subequations}
  \begin{equation}
H_0=J_1\sum_{[i,j]}S^z_iS^z_j+J_2\sum_{<i,j>}S^z_iS^z_j
  \end{equation}
\begin{equation}
H_1=J_1\sum_{[i,j]}(S^+_iS^-_j+S^-_iS^+_j)
\end{equation}
\begin{equation}
H_2=J_2\sum_{<i,j>}(S^+_iS^-_j+S^-_iS^+_j)
\end{equation}
\end{subequations}
and $\lambda$ is the expansion parameter. The limits $\lambda=0$ and $\lambda=1$ correspond to the Ising model and the isotropic Heisenberg 
model, respectively. $H_0$ is taken as the unperturbed Hamiltonian while $H'=H_1+H_2$ is the perturbation operator. To obtain a $T=0$ 
expansion about the NCAF phase, we rotate all the spins so as to have a ferromagnetic ground state. In this case the Hamiltonian form in (\ref{Hamiltonianform})
is still valid but now
\begin{subequations}
\begin{eqnarray}
H_0=J_1\cos(2q)\sum_{[i,j]}S^z_iS^z_j+\nonumber\\
+J_2\cos(q)\sum_{<i,j>}S^z_iS^z_j
\end{eqnarray}
\begin{eqnarray}
H_1=J_1\sum_{[i,j]}S^y_iS^y +\cos(2q)S^x_iS^x+\nonumber\\
+\sin(2q)(S^z_iS^x_j-S^x_iS^z_j)
\end{eqnarray}
\begin{eqnarray}
H_2=J_2\sum_{<i,j>}S^y_iS^y +\cos(q)S^x_iS^x+\nonumber\\
+\sin(q)(S^z_iS^x_j-S^x_iS^z_j)
\end{eqnarray}
\end{subequations}
where $q$ is the wave vector of the NCAF phase. Once the ground state phase has been chosen, perturbation theory 
can be applied, leading to an expansion of $H$ up to desired order in $\lambda$ for
the ground state wave function ($|\psi_{GS}>$), the ground state energy and other correlation functions.
The sublattice magnetization is given by,
\begin{eqnarray}
<M>=\frac{<\psi_{GS}|\hat{S}^z|\psi_{GS}>}{<\psi_{GS}|\psi_{GS}>}
\end{eqnarray}
The details behind the technique are discussed elsewhere and will not be repeated here. For a complete review see Ref.\cite{oitmaa2006,gelfand2000}

\section{One dimensional spin-half Heisenberg model ($J_2=0$)}\label{1Dspinhalf}
The ground state properties of the  1-D spin-half Heisenberg model at $T=0$ are well known from the exact solutions\cite{bethe1931}. 
It is also known\cite{baxter1973} 
that the ground state energy and sublattice magnetization,
 as a function of $\lambda$, have
essential singularities of the form
\begin{equation}\label{essential}
\exp(-{\frac{1}{\sqrt{1-\lambda}}})
\end{equation}
We first study the 1-D case, to see how well the series expansion methods can reproduce the exact results.
Series coefficients for the sublattice magnetization of this model are generated up to 
order $10$ in $\lambda$. 
Their anaylsis is carried out 
in two different ways: (i) using Integrated Differential Approximants (IDA) on the series obtained; (ii) using Biassed 
Integrated Differential Approximants (BIDA) on the natural logarithm of the same series. The first approach is the most straightforward and 
simply fits the known coefficients of the series to a homogeneous or inhomogeneous differential equation of the form
\begin{equation}\label{diffeq}
P_K(\lambda)\frac{d^2f}{d\lambda^2}+Q_L(\lambda)\frac{df}{d\lambda}+R_M(\lambda)f+S_T(\lambda)=0
\end{equation}
where $P_K,Q_L,R_M,S_T$ are polynomials of degree $K,L,M,T$ respectively. The results reported in this section are obtained by setting 
the polynomial $P_K$ and $S_T$ to zero, which is equivalent to a Dlog Pad\'{e} analysis
\begin{equation}\label{Dlogpade}
\frac{1}{f}\frac{df}{d\lambda}=-\frac{R_M(\lambda)}{Q_L(\lambda)}
\end{equation}
\begin{figure}[ht]
\begin{center}
\resizebox{80mm}{!}{\includegraphics{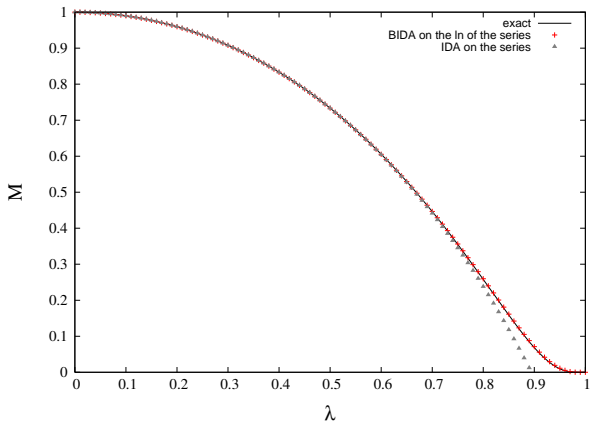}}
\caption{\label{fig:magnetization}(Color online) (a) Sublattice magnetization, normalized to unity in the Ising limit.
The red crosses represent the approximants obtained by analyzing the 
natural log of the series with BIDA. The grey triangles represent the approximants obtained by analyzing the original series by standard IDA}
\end{center}
\end{figure}
The second analysis method proceeds by taking the natural logarithm of the calculated series in order 
to reduce the essential singularity (\ref{essential}) into an algebraic one
\begin{equation}\label{algebraic}
\ln(e^{-\frac{1}{\sqrt{1-\lambda}}})=-\frac{1}{\sqrt{1-\lambda}}
\end{equation}
with critical exponents $\gamma =-0.5$. A sigularity of this form can easily be analyzed by using an IDA type of analysis. Moreover, a better 
convergence can be achieved by biassing the exponent in the analysis. This simply means that the function $f$ in (\ref{Dlogpade}) 
is forced to have an algebraic singularity of the form (\ref{algebraic}) with critical exponent $\gamma=-0.5$, as known from exact result. 
Fig.~\ref{fig:magnetization} shows the comparison between the two analysis methods and the exact result. 
While the two methods give very similar results for $\lambda < 0.7$, it is clear that they tend to disagree for $\lambda \rightarrow 1$.
The approximants obtained by Biassed IDA on the natural logarithm of the series, reproduces the behavior of the series close to the critical point much better.
 This is expected as it is extremely difficult to account for an essential singularity of the form (\ref{essential}) 
with a short, finite series without the biassing.
On the other hand, Fig.~\ref{fig:magnetization} shows that, once the essential singularity has been reduced to an algebraic one, by taking the natural logarithm of the series, 
even a finite series up to ten terms can reproduce quite well the exact behavior of the system all the way to the critical point.
\section{Spin-half Heisenberg model on anisotropic triangular lattice}\label{2Dspinhalf}
\begin{figure}[ht]
\begin{center}
\resizebox{80mm}{!}{\includegraphics{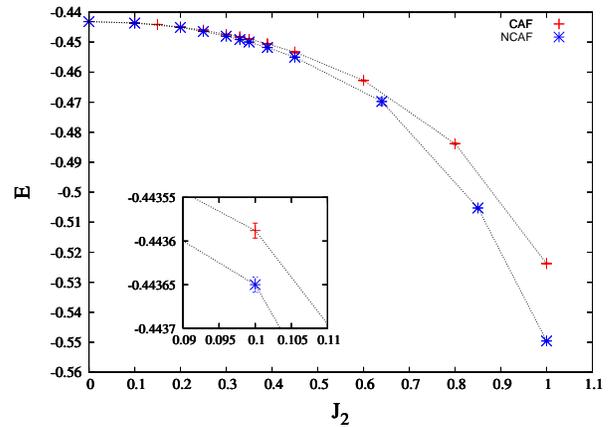}}
\caption{\label{fig:energy}(Color online) Energy for CAF (red cross) and NCAF (blue stars) phases. 
The inset shows a zoom in of the region around $J_2=0.1$.}
\end{center}
\end{figure}
The ground state energy for the Hamiltonian (1) has been computed up to order 10 in $\lambda$ for both
NCAF and CAF phases for different values of the interchain couplings $J_2$. For the NCAF phase, we consider a range of $q$-values and minimize the energy
with respect to $q$.
For each value of $J_2$, we then calculate the ratio series
\begin{equation}\label{ratio}
R_{J_2}(\lambda)=\frac{E_{J_2}(\lambda)}{E_{J_2=0}(\lambda)}
\end{equation}
where $E_{J_2}(\lambda)$ is the energy series calculated at a specific value of $J_2$ and $E_{J_2=0}(\lambda)$ is the energy series computed for the 1-D model. 
The idea behind this is that, if the series for $J_2\neq0$ has apparent singularities as a function of $\lambda$, as a consequence of being close to the 1-D limit,
we can eliminate its effect by taking the ratio. This allows us to evaluate the energy with increased accuracy.
Moreover, to improve convergence, an additional term is added to the Hamiltonian (2) as in previous studies\cite{oitmaa1996,Zheng1999}.
\begin{equation}
H=H_0+t\sum_iS^z_i+\lambda (H_1+H_2-t\sum_iS^z_i)
\end{equation}
For $\lambda \rightarrow 1$, the limit we are interested in, this has no effect on the Hamiltonian. The amplitude of the 
convergence term $t$ is generally taken as 1.
The extrapolated series ratio (\ref{ratio}) is multiplied by
the exact result for the 1-D case, $E=-\ln(2)+\frac{1}{4}$, to obtain the energies at different $J_2$.
Fig.~\ref{fig:energy} shows the obtained results. In this plot, the error bars reflect the spread of well-behaved approximants.
At $J_2=1$ the energy for the NCAF phase is centered around $-0.5508$, in
good agreement with earlier studies\cite{zheng2006}.
For $0<J_2<1$,  Fig.~\ref{fig:energy} shows that, in our calculations, the NCAF phase has a lower energy 
than the CAF phase for all values of $J_2$.
The inset shows that even for $J_2=0.1$, the lowest data point taken in our calculation, the NCAF phase appears to have lower energy than the CAF phase. 
However, while this result is suggestive that the NCAF phase is the
correct phase of the model, we should emphasize that our results are sensitive to how
the series are analyzed, especially near the decoupled chain limit.
Hence, further numerical studies of this model, looking in an unbiassed way at
short distance spin correlations, would be useful.

\subsection{Sublattice magnetization}
In this section, we study the sublattice magnetization for $J_2\neq 0$. 
Because the sublattice magnetization vanishes for the 1-D model,
it is not useful to consider the ratio of the 
sublattice magnetization series with that at $J_2=0$.

\begin{figure}[ht]
\begin{center}
\resizebox{75mm}{!}{\includegraphics{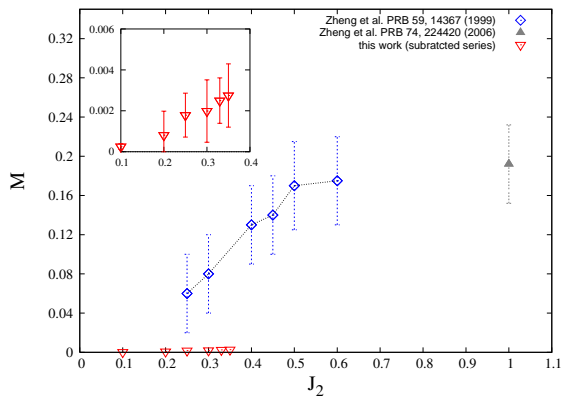}}
\caption{\label{fig:mag}(Color online) Sublattice magnetization for NCAF phase, normalized to a classical
value of $0.5$ calculated by Zheng et al.~\cite{Zheng1999,zheng2006} (blue squares and gray triangle) and in the present work after subtracting the 1D series (red triangles). See text for details. 
(inset) Zoom in of the small interchain coupling region showing the data points calculated in this work.} 
\end{center}
\end{figure}

We have calculated series expansion coefficients for the sublattice magnetization of the 
NCAF and CAF phases to order 10 in $\lambda$. 
Zheng et al.~\cite{Zheng1999,zheng2006} had earlier calculated
the sublattice magnetization for the classical spiral phase for
$0.25\leq J_2\leq 1$. These are shown in Fig.~\ref{fig:mag}. 
In the triangular lattice limit, the magnetization is $0.19(2)$, and as we move towards the 1-D limit, it begins decreasing
almost linearly to zero for $J_2\leq 0.5$. 
We have tried various approaches to analyzing the magnetization series. 
They lead to similar results but with no improved convergence.
The series analysis is even less reliable in the CAF phase. It is possible 
that the results at small $J_2$ are strongly influenced by the nearby essential singularity at $J_2=0$.

Motivated by the idea that the nearby singularity at $J_2=0$ may be influencing the series analysis,
we take the following approach: We subtract off for each value of $J_2\neq 0$ the series for the 1-D limit ($J_2=0$). 
This should eliminate the influence of 
the essential singularities. We then analyze the series with standard IDA. 
The results obtained in the region $0.1\leq J_2\leq 0.35$ 
following this procedure are shown in Fig.~\ref{fig:mag} (red triangles). This analysis leads to very small
values of the magnetization,
consistent with the exponentially small values expected from the
work of Bocquet et al.\cite{Bocquet2001}. Unfortunately this analysis does not smoothly
connect with the results at large $J_2$, so it is not clear how far in $J_2$ it should
be continued. If this scenario is correct, 
there maybe a sharp quantitative change of behavior between small
and large $J_2$. We note that other
groups have even suggested various phase transitions
as a function of $J_2$\cite{Weng2006,yunoki2006}.
A similar analysis for the sublattice magnetization 
series for the CAF phase gives only negative values. 
\section{spin one}\label{spinone}
\begin{figure}[ht]\label{wavevector}
\resizebox{70mm}{!}{\includegraphics{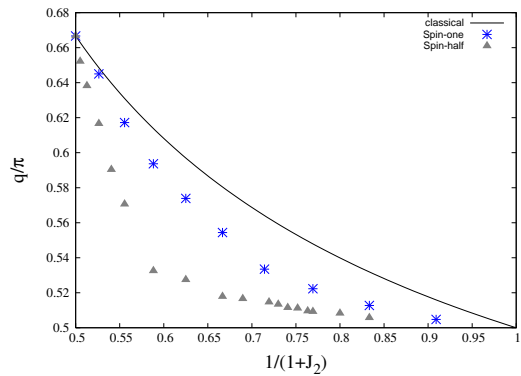}}
\caption{\label{fig:wavevector}(Color online) Comparison between the NCAF phase wave vector $q$ (in units of $\pi$) for the spin-1 and spin-$\frac{1}{2}$ 2-D Heisenberg 
model on anisotropic triangular lattice as a function of $J_2$. The classical result is also shown by a black solid line.}
\end{figure}

In this section we present calculations for the ground state properties of the spin-one model
on the anisotropic triangular-lattice. Ground state energy and sublattice magnetization have
been calculated for CAF and NCAF phases as for the spin-half model.
We begin by showing in
Fig.~\ref{fig:wavevector} the wave vector $q$, in units of $\pi$, for the non colinear antiferromagnetic phase (NCAF) as a function of $J_2$. 
For each value of $J_2$, the energy has been minimized with respect to $q$ and the value of $q$ for which $E=E_{min}$ has been plotted. 
The analysis has been carried out 
for the spin-$\frac{1}{2}$ and the spin-1 Heisenberg model. The classical result $q=\cos^{-1}(-\frac{J_2}{2J_1})$ 
is shown by a solid black line.
The NCAF phase for the spin-1 system is closer to the classical solution than the spin-$\frac{1}{2}$ system, as expected. 
Nevertheless, evidence for quantum effects in the properties of the ground state are clearly evident.

\subsection{Phase Diagram for anisotropic square and triangular lattice spin-one models}
In the 1-D limit, it is well known that the Heisenberg system is in the Haldane gap phase\cite{affleck1989a}.
Furthermore, Ising expansions lead to a critical point before the Heisenberg
symmetry is restored ($\lambda_c<1$)\cite{singh1988}.
Once, the couplings $J_2$ are turned on, we would like to follow $\lambda_c$ as
a function of $J_2$ to see when it reaches unity. This will tell us the critical $J_2$
required to close the Haldane gap for the Heisenberg system.
\begin{figure}[ht]\label{lambdacrit}
\resizebox{70mm}{!}{\includegraphics{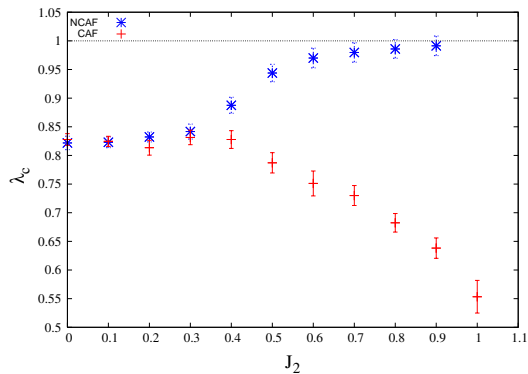}}
\caption{\label{fig:lambdacrit} (Color online) $\lambda_c$ vs. interchain coupling $J_2$ for NCAF and CAF phase of the spin-1 Heisenberg model on the triangular 
lattice.} 
\end{figure}

\begin{figure}[ht]\label{m}
\resizebox{70mm}{!}{\includegraphics{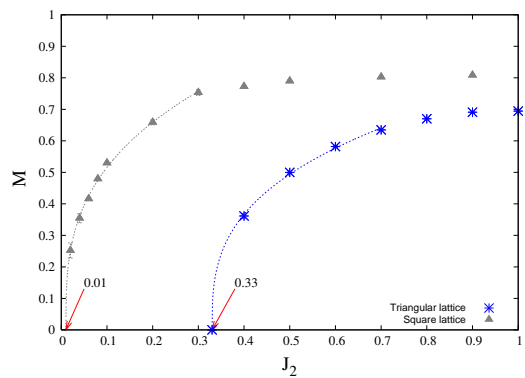}}
\caption{\label{fig:m}(Color online) (blue stars) Magnetization vs. interchain coupling $J_2$ for the NCAF phase of the spin-1 Heisenberg model on the triangular lattice. 
(grey triangles) Magnetization vs. interchain coupling $J_2$ for the Ne\'{e}l phase of the spin-1 Heisenberg model on the square lattice.
(dotted lines) Fit of the magnetization curve to a power law of the form given in equation~(\ref{powerlaw})}
\end{figure}
In Fig.~\ref{fig:lambdacrit}, we show the value of the anisotropy parameter $\lambda_c$ 
as a function of the couplings $J_2$ obtained by a Dlog Pad\'{e} analysis of the sublattice magnetization series.
One can see a clear difference between the NCAF and CAF phases. The CAF phase becomes less and less stable
with interchain coupling and is never realized for the Heisenberg models ($\lambda$ equal to unity).
In contrast, the spiral phase become more stable with interchain coupling and for some $J_2$ the critical $\lambda$
reaches unity. It turns out that this way of studying the critical $J_2$ is less accurate than an
alternative method discussed below.

A better way to compute critical $J_2$ needed to close the Haldane gap is to consider the
sublattice magnetization as a function of $J_2$ coming from the triangular-lattice side (large $J_2$).
The series analysis should be most accurate when the system is well ordered.
In Fig.~\ref{fig:m}, we show the calculated sublattice magnetization
for the Heisenberg model for both the square-lattice type unfrustrated interchain couplings
and the triangular-lattice type frustrated interchain couplings. The sublattice magnetization as a function of $J_2$ 
is then fitted to a power law of the form
\begin{equation}\label{powerlaw}
M(J_2)=(J_2-J_2^{crit})^{\gamma}
\end{equation}
in which $J_2^{crit}$ and $\gamma$ are free parameters. This leads to estimates 
for $J_2^{crit}$ of approximately $0.01$ and $0.33$ respectively with exponent values which are around a third, as expected for 3-D models\cite{ma1976}. 
The value of $J_2^{crit}\approx 0.01$ for the former is consistent with chain mean-field theory estimates\cite{Affleck1989}.
It is clear that this kind of frustration significantly extends the Haldane gap phase. Note that this is very different from a material like CsNiCl$_3$,\cite{Buyers1986} where a given chain has six neighboring
chains arranged in a triangular-lattice geometry. In that case, the chains themselves remain unfrustrated
and frustration only affects the relative spin-orientation between the chains\cite{Affleck1989}. The latter is much less effective in preventing long-range order.

\section{Conclusions}\label{conclusions}
In this paper, we have studied the spin-half and spin-one Heisenberg models in spin-chains that are coupled in an 
anisotropic triangular-lattice geometry, where a spin in one chain is coupled to two neighboring spins in another chain. 
This geometry is particularly effective in preventing the development of spin
correlations between chains, and in altering spin correlations within a chain. 
For the spin-half case, both Colinear Antiferromagnetic phase and Non-colinear Antiferromagnetic
phase are possible in this geometry. Though, in our calculations, the Non-colinear phase appears to have a lower energy.
Further unbiassed ways of studying this competition numerically would be useful. Short distance spin correlations can be used to distinguish between the different phases.
Though, potential biassing due to boundary conditions should be taken into account.
For the spin-one model also, we find that
this geometry significantly enhances the stability of one-dimensional Haldane gap phase,
and prevents the development of long range magnetic order. 
This geometry is quite different from materials like CsNiCl$_3$,
where frustration does not lead to competing correlations along any given chain and thus the Haldane gap phase is
quickly destroyed.
It would be interesting if materials similar to the spin-half materials Cs$_2$CuCl$_4$ and Cs$_2$CuBr$_4$ are found with spin-one.
The study of such materials would shed further light on the role of interchain frustration and the extension of one-dimensional
physics to higher dimensional systems.

We would like to thank Ross Mckenzie for discussions. 
This work is supported in part by the National Science Foundation under grant numbers
DMR-0240918 and PHY05-51164.


\begin{thebibliography}{39}
\expandafter\ifx\csname natexlab\endcsname\relax\def\natexlab#1{#1}\fi
\expandafter\ifx\csname bibnamefont\endcsname\relax
  \def\bibnamefont#1{#1}\fi
\expandafter\ifx\csname bibfnamefont\endcsname\relax
  \def\bibfnamefont#1{#1}\fi
\expandafter\ifx\csname citenamefont\endcsname\relax
  \def\citenamefont#1{#1}\fi
\expandafter\ifx\csname url\endcsname\relax
  \def\url#1{\texttt{#1}}\fi
\expandafter\ifx\csname urlprefix\endcsname\relax\def\urlprefix{URL }\fi
\providecommand{\bibinfo}[2]{#2}
\providecommand{\eprint}[2][]{\url{#2}}

\bibitem[{\citenamefont{Auerbach}(1994)}]{Auerbach1994}
\bibinfo{author}{\bibfnamefont{A.}~\bibnamefont{Auerbach}},
  \emph{\bibinfo{title}{Interacting Electrons and Quantum Magnetism}}
  (\bibinfo{publisher}{Springer-Verlag, New York}, \bibinfo{year}{1994}).

\bibitem[{\citenamefont{Sachdev}(1995)}]{Sachdev1995}
\bibinfo{author}{\bibfnamefont{S.}~\bibnamefont{Sachdev}},
  \emph{\bibinfo{title}{Low Dimensional Quantum Field Theories for Condensed
  Matter Physicists}} (\bibinfo{publisher}{World Scientific, Singapore},
  \bibinfo{year}{1995}), \bibinfo{note}{edited by Y. Lu, S. Lundqvist, and G.
  Morandi}.

\bibitem[{\citenamefont{Chakravarty et~al.}(1988)\citenamefont{Chakravarty,
  Halperin, and Nelson}}]{Chakravarty1988}
\bibinfo{author}{\bibfnamefont{S.}~\bibnamefont{Chakravarty}},
  \bibinfo{author}{\bibfnamefont{B.~I.} \bibnamefont{Halperin}},
  \bibnamefont{and} \bibinfo{author}{\bibfnamefont{D.~R.}
  \bibnamefont{Nelson}}, \bibinfo{journal}{Phys. Rev. Lett.}
  \textbf{\bibinfo{volume}{60}}, \bibinfo{pages}{1057} (\bibinfo{year}{1988}).

\bibitem[{\citenamefont{Zheng et~al.}(1999)\citenamefont{Zheng, McKenzie, and
  Singh}}]{Zheng1999}
\bibinfo{author}{\bibfnamefont{W.}~\bibnamefont{Zheng}},
  \bibinfo{author}{\bibfnamefont{R.~H.} \bibnamefont{McKenzie}},
  \bibnamefont{and} \bibinfo{author}{\bibfnamefont{R.~R.~P.}
  \bibnamefont{Singh}}, \bibinfo{journal}{Phys. Rev. B}
  \textbf{\bibinfo{volume}{59}}, \bibinfo{pages}{14367} (\bibinfo{year}{1999}).

\bibitem[{\citenamefont{Bernu et~al.}(1994)\citenamefont{Bernu, Lecheminant,
  Lhuillier, and Pierre}}]{Bernu1994}
\bibinfo{author}{\bibfnamefont{B.}~\bibnamefont{Bernu}},
  \bibinfo{author}{\bibfnamefont{P.}~\bibnamefont{Lecheminant}},
  \bibinfo{author}{\bibfnamefont{C.}~\bibnamefont{Lhuillier}},
  \bibnamefont{and} \bibinfo{author}{\bibfnamefont{L.}~\bibnamefont{Pierre}},
  \bibinfo{journal}{Phys. Rev. B} \textbf{\bibinfo{volume}{50}},
  \bibinfo{pages}{10048} (\bibinfo{year}{1994}).

\bibitem[{\citenamefont{Singh and Huse}(1992)}]{Singh1992}
\bibinfo{author}{\bibfnamefont{R.~R.~P.} \bibnamefont{Singh}} \bibnamefont{and}
  \bibinfo{author}{\bibfnamefont{D.~A.} \bibnamefont{Huse}},
  \bibinfo{journal}{Phys. Rev. Lett.} \textbf{\bibinfo{volume}{68}},
  \bibinfo{pages}{1766} (\bibinfo{year}{1992}).

\bibitem[{\citenamefont{Zhong and Sorella}(1993)}]{Zhong1993}
\bibinfo{author}{\bibfnamefont{Q.~F.} \bibnamefont{Zhong}} \bibnamefont{and}
  \bibinfo{author}{\bibfnamefont{S.}~\bibnamefont{Sorella}},
  \bibinfo{journal}{Europhys. Lett.} \textbf{\bibinfo{volume}{21}},
  \bibinfo{pages}{629} (\bibinfo{year}{1993}).

\bibitem[{\citenamefont{Capriotti et~al.}(1999)\citenamefont{Capriotti,
  Trumper, and Sorella}}]{capriotti1999}
\bibinfo{author}{\bibfnamefont{L.}~\bibnamefont{Capriotti}},
  \bibinfo{author}{\bibfnamefont{A.~E.} \bibnamefont{Trumper}},
  \bibnamefont{and} \bibinfo{author}{\bibfnamefont{S.}~\bibnamefont{Sorella}},
  \bibinfo{journal}{Phys. Rev. Lett.} \textbf{\bibinfo{volume}{82}},
  \bibinfo{pages}{3899} (\bibinfo{year}{1999}).

\bibitem[{\citenamefont{Coldea et~al.}(1997)\citenamefont{Coldea, Tennant,
  Cowley, McMorrow, Dorner, and Tylczynski}}]{Coldea1997}
\bibinfo{author}{\bibfnamefont{R.}~\bibnamefont{Coldea}},
  \bibinfo{author}{\bibfnamefont{D.~A.} \bibnamefont{Tennant}},
  \bibinfo{author}{\bibfnamefont{R.~A.} \bibnamefont{Cowley}},
  \bibinfo{author}{\bibfnamefont{D.~F.} \bibnamefont{McMorrow}},
  \bibinfo{author}{\bibfnamefont{B.}~\bibnamefont{Dorner}}, \bibnamefont{and}
  \bibinfo{author}{\bibfnamefont{Z.}~\bibnamefont{Tylczynski}},
  \bibinfo{journal}{Phys. Rev. Lett.} \textbf{\bibinfo{volume}{79}},
  \bibinfo{pages}{151} (\bibinfo{year}{1997}).

\bibitem[{\citenamefont{Coldea et~al.}(2002)\citenamefont{Coldea, Tennant,
  Habicht, Smeibidl, Wolters, and Tylczynski}}]{Coldea2002}
\bibinfo{author}{\bibfnamefont{R.}~\bibnamefont{Coldea}},
  \bibinfo{author}{\bibfnamefont{D.~A.} \bibnamefont{Tennant}},
  \bibinfo{author}{\bibfnamefont{K.}~\bibnamefont{Habicht}},
  \bibinfo{author}{\bibfnamefont{P.}~\bibnamefont{Smeibidl}},
  \bibinfo{author}{\bibfnamefont{C.}~\bibnamefont{Wolters}}, \bibnamefont{and}
  \bibinfo{author}{\bibfnamefont{Z.}~\bibnamefont{Tylczynski}},
  \bibinfo{journal}{Phys. Rev. Lett.} \textbf{\bibinfo{volume}{88}},
  \bibinfo{pages}{137203} (\bibinfo{year}{2002}).

\bibitem[{\citenamefont{Coldea et~al.}(2003)\citenamefont{Coldea, Tennant, and
  Tylczynski}}]{Coldea2003}
\bibinfo{author}{\bibfnamefont{R.}~\bibnamefont{Coldea}},
  \bibinfo{author}{\bibfnamefont{D.~A.} \bibnamefont{Tennant}},
  \bibnamefont{and}
  \bibinfo{author}{\bibfnamefont{Z.}~\bibnamefont{Tylczynski}},
  \bibinfo{journal}{Phys. Rev. B} \textbf{\bibinfo{volume}{68}},
  \bibinfo{pages}{134424} (\bibinfo{year}{2003}).

\bibitem[{\citenamefont{Coldea et~al.}(1996)\citenamefont{Coldea, Tennant,
  Cowley, McMorrow, Dorner, and Tylczynski}}]{Coldea1996b}
\bibinfo{author}{\bibfnamefont{R.}~\bibnamefont{Coldea}},
  \bibinfo{author}{\bibfnamefont{D.~A.} \bibnamefont{Tennant}},
  \bibinfo{author}{\bibfnamefont{R.~A.} \bibnamefont{Cowley}},
  \bibinfo{author}{\bibfnamefont{D.~F.} \bibnamefont{McMorrow}},
  \bibinfo{author}{\bibfnamefont{B.}~\bibnamefont{Dorner}}, \bibnamefont{and}
  \bibinfo{author}{\bibfnamefont{Z.}~\bibnamefont{Tylczynski}},
  \bibinfo{journal}{J. Phys.: Cond. Matt.} \textbf{\bibinfo{volume}{8}},
  \bibinfo{pages}{7473} (\bibinfo{year}{1996}).

\bibitem[{\citenamefont{Veillette et~al.}(2005)\citenamefont{Veillette, James,
  and Essler}}]{veillette2005}
\bibinfo{author}{\bibfnamefont{M.~Y.} \bibnamefont{Veillette}},
  \bibinfo{author}{\bibfnamefont{A.~J.~A.} \bibnamefont{James}},
  \bibnamefont{and} \bibinfo{author}{\bibfnamefont{F.~H.~L.}
  \bibnamefont{Essler}}, \bibinfo{journal}{Phys. Rev. B}
  \textbf{\bibinfo{volume}{72}}, \bibinfo{eid}{134429} (\bibinfo{year}{2005}).

\bibitem[{\citenamefont{Dalidovich et~al.}(2006)\citenamefont{Dalidovich,
  Sknepnek, Berlinsky, Zhang, and Kallin}}]{dalidovich2006}
\bibinfo{author}{\bibfnamefont{D.}~\bibnamefont{Dalidovich}},
  \bibinfo{author}{\bibfnamefont{R.}~\bibnamefont{Sknepnek}},
  \bibinfo{author}{\bibfnamefont{A.~J.} \bibnamefont{Berlinsky}},
  \bibinfo{author}{\bibfnamefont{J.}~\bibnamefont{Zhang}}, \bibnamefont{and}
  \bibinfo{author}{\bibfnamefont{C.}~\bibnamefont{Kallin}},
  \bibinfo{journal}{Phys. Rev. B} \textbf{\bibinfo{volume}{73}},
  \bibinfo{eid}{184403} (\bibinfo{year}{2006}).

\bibitem[{\citenamefont{Fjaerestad et~al.}(2007)\citenamefont{Fjaerestad,
  Zheng, Singh, McKenzie, and Coldea}}]{restad2007}
\bibinfo{author}{\bibfnamefont{J.~O.} \bibnamefont{Fjaerestad}},
  \bibinfo{author}{\bibfnamefont{W.}~\bibnamefont{Zheng}},
  \bibinfo{author}{\bibfnamefont{R.~R.~P.} \bibnamefont{Singh}},
  \bibinfo{author}{\bibfnamefont{R.~H.} \bibnamefont{McKenzie}},
  \bibnamefont{and} \bibinfo{author}{\bibfnamefont{R.}~\bibnamefont{Coldea}},
  \bibinfo{journal}{Phys. Rev. B} \textbf{\bibinfo{volume}{75}},
  \bibinfo{eid}{174447} (\bibinfo{year}{2007}).

\bibitem[{zhe()}]{zheng2006}
\bibinfo{note}{W. Zheng, J. O. Fjaerestad, R. R. P. Singh, R. H. McKenzie and
  R. Coldea, Phys. Rev. B $\bf{74}$, 224420 (2006); W. Zheng, J. O. Fjaerestad,
  R. R. P. Singh, R. H. McKenzie and R. Coldea, Phys. Rev. Lett. $\bf{96}$,
  057201, (2001).}

\bibitem[{\citenamefont{Zheng et~al.}(2005)\citenamefont{Zheng, Singh,
  McKenzie, and Coldea}}]{zheng2005}
\bibinfo{author}{\bibfnamefont{W.}~\bibnamefont{Zheng}},
  \bibinfo{author}{\bibfnamefont{R.~R.~P.} \bibnamefont{Singh}},
  \bibinfo{author}{\bibfnamefont{R.~H.} \bibnamefont{McKenzie}},
  \bibnamefont{and} \bibinfo{author}{\bibfnamefont{R.}~\bibnamefont{Coldea}},
  \bibinfo{journal}{Phys. Rev. B} \textbf{\bibinfo{volume}{71}},
  \bibinfo{eid}{134422} (\bibinfo{year}{2005}).

\bibitem[{\citenamefont{Kohno et~al.}(2007)\citenamefont{Kohno, Starykh, and
  Balents}}]{Kohno2007}
\bibinfo{author}{\bibfnamefont{M.}~\bibnamefont{Kohno}},
  \bibinfo{author}{\bibfnamefont{O.~A.} \bibnamefont{Starykh}},
  \bibnamefont{and} \bibinfo{author}{\bibfnamefont{L.}~\bibnamefont{Balents}},
  \bibinfo{journal}{Nature Physics} \textbf{\bibinfo{volume}{3}},
  \bibinfo{pages}{790} (\bibinfo{year}{2007}).

\bibitem[{\citenamefont{Starykh and Balents}(2007)}]{starykh2007}
\bibinfo{author}{\bibfnamefont{O.~A.} \bibnamefont{Starykh}} \bibnamefont{and}
  \bibinfo{author}{\bibfnamefont{L.}~\bibnamefont{Balents}},
  \bibinfo{journal}{Phys. Rev. Lett.} \textbf{\bibinfo{volume}{98}},
  \bibinfo{eid}{077205} (\bibinfo{year}{2007}).

\bibitem[{\citenamefont{Schulz et~al.}(1996)\citenamefont{Schulz, Ziman, and
  Poilblanc}}]{schulz}
\bibinfo{author}{\bibfnamefont{H.~J.} \bibnamefont{Schulz}},
  \bibinfo{author}{\bibfnamefont{T.~A.~L.} \bibnamefont{Ziman}},
  \bibnamefont{and}
  \bibinfo{author}{\bibfnamefont{D.}~\bibnamefont{Poilblanc}},
  \bibinfo{journal}{J. Phys. I France} \textbf{\bibinfo{volume}{6}},
  \bibinfo{pages}{675} (\bibinfo{year}{1996}).

\bibitem[{\citenamefont{Oitmaa and Zheng}(1996)}]{oitmaa1996}
\bibinfo{author}{\bibfnamefont{J.}~\bibnamefont{Oitmaa}} \bibnamefont{and}
  \bibinfo{author}{\bibfnamefont{W.}~\bibnamefont{Zheng}},
  \bibinfo{journal}{Phys. Rev. B} \textbf{\bibinfo{volume}{54}},
  \bibinfo{pages}{3022} (\bibinfo{year}{1996}).

\bibitem[{\citenamefont{Alicea et~al.}(2005)\citenamefont{Alicea, Motrunich,
  and Fisher}}]{alicea2005}
\bibinfo{author}{\bibfnamefont{J.}~\bibnamefont{Alicea}},
  \bibinfo{author}{\bibfnamefont{O.~I.} \bibnamefont{Motrunich}},
  \bibnamefont{and} \bibinfo{author}{\bibfnamefont{M.~P.~A.}
  \bibnamefont{Fisher}}, \bibinfo{journal}{Phys. Rev. Lett.}
  \textbf{\bibinfo{volume}{95}}, \bibinfo{eid}{247203} (\bibinfo{year}{2005}).

\bibitem[{\citenamefont{Alicea et~al.}(2006)\citenamefont{Alicea, Motrunich,
  and Fisher}}]{alicea2006}
\bibinfo{author}{\bibfnamefont{J.}~\bibnamefont{Alicea}},
  \bibinfo{author}{\bibfnamefont{O.~I.} \bibnamefont{Motrunich}},
  \bibnamefont{and} \bibinfo{author}{\bibfnamefont{M.~P.~A.}
  \bibnamefont{Fisher}}, \bibinfo{journal}{Phys. Rev. B}
  \textbf{\bibinfo{volume}{73}}, \bibinfo{eid}{174430} (\bibinfo{year}{2006}).

\bibitem[{\citenamefont{Yunoki and Sorella}(2006)}]{yunoki2006}
\bibinfo{author}{\bibfnamefont{S.}~\bibnamefont{Yunoki}} \bibnamefont{and}
  \bibinfo{author}{\bibfnamefont{S.}~\bibnamefont{Sorella}},
  \bibinfo{journal}{Phys. Rev. B} \textbf{\bibinfo{volume}{74}},
  \bibinfo{eid}{014408} (\bibinfo{year}{2006}).

\bibitem[{\citenamefont{Chung et~al.}(2001)\citenamefont{Chung, Marston, and
  McKenzie}}]{Chung2001}
\bibinfo{author}{\bibfnamefont{C.~H.} \bibnamefont{Chung}},
  \bibinfo{author}{\bibfnamefont{J.~B.} \bibnamefont{Marston}},
  \bibnamefont{and} \bibinfo{author}{\bibfnamefont{R.~H.}
  \bibnamefont{McKenzie}}, \bibinfo{journal}{Journal of Physics: Condensed
  Matter} \textbf{\bibinfo{volume}{13}}, \bibinfo{pages}{5159}
  (\bibinfo{year}{2001}).

\bibitem[{\citenamefont{Merino et~al.}(1999)\citenamefont{Merino, McKenzie,
  Marston, and Chung}}]{Merino1999}
\bibinfo{author}{\bibfnamefont{J.}~\bibnamefont{Merino}},
  \bibinfo{author}{\bibfnamefont{R.~H.} \bibnamefont{McKenzie}},
  \bibinfo{author}{\bibfnamefont{J.~B.} \bibnamefont{Marston}},
  \bibnamefont{and} \bibinfo{author}{\bibfnamefont{C.~H.} \bibnamefont{Chung}},
  \bibinfo{journal}{Journal of Physics: Condensed Matter}
  \textbf{\bibinfo{volume}{11}}, \bibinfo{pages}{2965} (\bibinfo{year}{1999}).

\bibitem[{\citenamefont{Weng et~al.}(2006)\citenamefont{Weng, Sheng, Weng, and
  Bursill}}]{Weng2006}
\bibinfo{author}{\bibfnamefont{M.~Q.} \bibnamefont{Weng}},
  \bibinfo{author}{\bibfnamefont{D.~N.} \bibnamefont{Sheng}},
  \bibinfo{author}{\bibfnamefont{Z.~Y.} \bibnamefont{Weng}}, \bibnamefont{and}
  \bibinfo{author}{\bibfnamefont{R.~J.} \bibnamefont{Bursill}},
  \bibinfo{journal}{Phys. Rev. B} \textbf{\bibinfo{volume}{74}},
  \bibinfo{eid}{012407} (\bibinfo{year}{2006}).

\bibitem[{\citenamefont{Moukouri}(2004)}]{moukouri2004}
\bibinfo{author}{\bibfnamefont{S.}~\bibnamefont{Moukouri}},
  \bibinfo{journal}{Phys. Rev. B} \textbf{\bibinfo{volume}{70}},
  \bibinfo{pages}{014403} (\bibinfo{year}{2004}).

\bibitem[{\citenamefont{Arlego and Brenig}(2007)}]{arlego2007}
\bibinfo{author}{\bibfnamefont{M.}~\bibnamefont{Arlego}} \bibnamefont{and}
  \bibinfo{author}{\bibfnamefont{W.}~\bibnamefont{Brenig}},
  \bibinfo{journal}{Phys. Rev. B} \textbf{\bibinfo{volume}{75}},
  \bibinfo{eid}{024409} (\bibinfo{year}{2007}).

\bibitem[{\citenamefont{Gelfand and Singh}(2000)}]{gelfand2000}
\bibinfo{author}{\bibfnamefont{M.~P.} \bibnamefont{Gelfand}} \bibnamefont{and}
  \bibinfo{author}{\bibfnamefont{R.~R.} \bibnamefont{Singh}},
  \bibinfo{journal}{Adv. Phys.} \textbf{\bibinfo{volume}{49}},
  \bibinfo{pages}{93} (\bibinfo{year}{2000}).

\bibitem[{\citenamefont{Bethe}(1931)}]{bethe1931}
\bibinfo{author}{\bibfnamefont{H.}~\bibnamefont{Bethe}}, \bibinfo{journal}{Z.
  Phys.} \textbf{\bibinfo{volume}{71}}, \bibinfo{pages}{205}
  (\bibinfo{year}{1931}).

\bibitem[{\citenamefont{Baxter}(1973)}]{baxter1973}
\bibinfo{author}{\bibfnamefont{R.~J.} \bibnamefont{Baxter}},
  \bibinfo{journal}{J. Stat. Phys.} \textbf{\bibinfo{volume}{9}},
  \bibinfo{pages}{145} (\bibinfo{year}{1973}).

\bibitem[{\citenamefont{Oitmaa et~al.}(2006)\citenamefont{Oitmaa, Hamer, and
  Zheng}}]{oitmaa2006}
\bibinfo{author}{\bibfnamefont{J.}~\bibnamefont{Oitmaa}},
  \bibinfo{author}{\bibfnamefont{C.}~\bibnamefont{Hamer}}, \bibnamefont{and}
  \bibinfo{author}{\bibfnamefont{W.}~\bibnamefont{Zheng}},
  \emph{\bibinfo{title}{Series Expansion Methods for Strongly Interacting
  Lattice Models}} (\bibinfo{publisher}{Cambridge University Press, New York},
  \bibinfo{year}{2006}).

\bibitem[{\citenamefont{Bocquet et~al.}(2001)\citenamefont{Bocquet, Essler,
  Tsvelik, and Gogolin}}]{Bocquet2001}
\bibinfo{author}{\bibfnamefont{M.}~\bibnamefont{Bocquet}},
  \bibinfo{author}{\bibfnamefont{F.~H.~L.} \bibnamefont{Essler}},
  \bibinfo{author}{\bibfnamefont{A.~M.} \bibnamefont{Tsvelik}},
  \bibnamefont{and} \bibinfo{author}{\bibfnamefont{A.~O.}
  \bibnamefont{Gogolin}}, \bibinfo{journal}{Phys. Rev. B}
  \textbf{\bibinfo{volume}{64}}, \bibinfo{pages}{094425}
  (\bibinfo{year}{2001}).

\bibitem[{\citenamefont{Affleck}(1989{\natexlab{a}})}]{affleck1989a}
\bibinfo{author}{\bibfnamefont{I.}~\bibnamefont{Affleck}}, \bibinfo{journal}{J.
  Phys.: Cond. Matt.} \textbf{\bibinfo{volume}{1}}, \bibinfo{pages}{3047}
  (\bibinfo{year}{1989}{\natexlab{a}}).

\bibitem[{\citenamefont{Singh and Gelfand}(1988)}]{singh1988}
\bibinfo{author}{\bibfnamefont{R.~R.~P.} \bibnamefont{Singh}} \bibnamefont{and}
  \bibinfo{author}{\bibfnamefont{M.~P.} \bibnamefont{Gelfand}},
  \bibinfo{journal}{Phys. Rev. Lett.} \textbf{\bibinfo{volume}{61}},
  \bibinfo{pages}{2133} (\bibinfo{year}{1988}).

\bibitem[{\citenamefont{Ma}(1976)}]{ma1976}
\bibinfo{author}{\bibfnamefont{S.~K.} \bibnamefont{Ma}},
  \emph{\bibinfo{title}{Modern Theory of Critical Phenomena}}
  (\bibinfo{publisher}{Benjamin Cummings, London}, \bibinfo{year}{1976}).

\bibitem[{\citenamefont{Affleck}(1989{\natexlab{b}})}]{Affleck1989}
\bibinfo{author}{\bibfnamefont{I.}~\bibnamefont{Affleck}},
  \bibinfo{journal}{Phys. Rev. Lett.} \textbf{\bibinfo{volume}{62}},
  \bibinfo{pages}{474} (\bibinfo{year}{1989}{\natexlab{b}}).

\bibitem[{\citenamefont{Buyers et~al.}(1986)\citenamefont{Buyers, Morra,
  Armstrong, Hogan, Gerlach, and Hirakawa}}]{Buyers1986}
\bibinfo{author}{\bibfnamefont{W.~J.~L.} \bibnamefont{Buyers}},
  \bibinfo{author}{\bibfnamefont{R.~M.} \bibnamefont{Morra}},
  \bibinfo{author}{\bibfnamefont{R.~L.} \bibnamefont{Armstrong}},
  \bibinfo{author}{\bibfnamefont{M.~J.} \bibnamefont{Hogan}},
  \bibinfo{author}{\bibfnamefont{P.}~\bibnamefont{Gerlach}}, \bibnamefont{and}
  \bibinfo{author}{\bibfnamefont{K.}~\bibnamefont{Hirakawa}},
  \bibinfo{journal}{Phys. Rev. Lett.} \textbf{\bibinfo{volume}{56}},
  \bibinfo{pages}{371} (\bibinfo{year}{1986}).

\end{thebibliography}
\end{document}